# Parametric Estimation of Handoff

*Soumen Kanrar*
*Vehere interactive Pvt Ltd*
Calcutta India
Soumen.kanrar@veheretech.com

*Abstract*— The efficiency of wireless technology depends upon the seamless connectivity to the user at anywhere any time. Heterogeneous wireless networks are an integration of different networks with diversified technologies. The most essential requirement for Seamless vertical handover is that the received signal strength should always be healthy. Mobile device enabled with multiple wireless technologies makes it possible to maintain seamless connectivity in highly dynamic environment. Since the available bandwidth is limited and the number of users is growing rapidly, it's a real challenge to maintain the received signal strength in a healthy stage. In this work, the proposed, cost-effective parametric estimation for vertical handover shows that the received signal strength maintains a healthy level by considering the novel concept.

*Keywords-Vertical handover, heterogeneous network, received signal strength, cost function, parametric estimation.*

## I INTRODUCTION

The next-generation wireless networks are heterogeneous by nature. In the heterogeneous behavior, the radio resource management is one of the most important issues. Theoretically, the bandwidth allocation to the different cells which cover a total zone is uniformly distributed. One of the key issues in the bandwidth allocation is the frequency reusability among the cells in that zone (considering macro micro and pico cell). The call admission control is used to enhance the system performance. In call admission control, two parameters play major role in the system performance. One is call blocking probability in the wireless networks. The call blocking probability is one of such quality of service parameters for the wireless network. So for the better quality of service, it is desirable to reduce the call blocking probability by optimizing the utilization of available radio resources. Another important parameter is Handoff blocking and call dropping probability. The first determines the fraction of new calls that are blocked, while the second is closely related to the fraction of admitted calls that terminate prematurely due to dropout. Handoff calls to the neighboring cells should be considered with higher priority than the new submit call in that cell. So the handoff call must consider immediately. This is because an abrupt premature termination of an ongoing conversation will upset the caller more than rejection of the call in the first place. The optimization problem is to reduce or minimize the handoff drop probability when call approaching from nearby neighbor cell. One way to solve this problem that every base station has to reserve some amount of bandwidth. The reserve bandwidth to be serving to the handoff call on an emergency basis. The channel allocation is done to the new call from the remaining spectrum of that particular cell. The traffic load is not uniformly distributed to each cell and in the real scenario, some numbers of cells are highly dense and some are less. The major problems are remaining like the desirable strategies for the bandwidth distribution among the cells such that the optimum size of the reserve bandwidth for each cell. In every case, the goal is to reduce the Handoff dropping or call blocking. Here we address the problem in a different way by considering the received signal strength, the covering zone where the received signal strength above some threshold values and the latency. In late nineties' Stemmand Katz [1] focus upon the area of vertical hand over for seamless connectivity in heterogeneous network. During that time, the author considered the available technology like wireless LAN, Infrared wireless LAN, Wave LAN etc. The problem of call blocking and handoff call dropping probabilities can be handling by acceptable tradeoff between those attributes [9]. In the recent works focused on the Low-latency handoff schemes used to analysis the SIP based mobility management and Mobility management for VoIP in 3G systems [2, 3, 4]. Traditional soft handover polices [11, 12] are based on Received Signal Strength (RSS) or its relative equivalent Received Signal Strength Indicator (RSSI) comparisons. In the homogeneous network, RSS policies may be sufficient, but in heterogeneous network these policies are not reliable and correct. In this work proposed a cost effective parametric estimation to make the soft vertical handover smooth with reducing the call dropping probabilities.

## II PROBLEM DESCRIPTION

In this model, the user terminal or mobile node moves inside the hexagonal cell, the cell, itself has large coverage area and limited bandwidth. Inside the cell numbers of wireless technology exist. Assuming the shape of the service domain are heterogeneous by nature Fig 1. The technologies provide to the user .i.e. mobile node as narrow band and broadband services. The narrow band basically use for the voice channel and the broad band used in the multimedia service. We assume that the mobile node moves in a nonlinear path. For the seamless vertical handover, the cost functions to be optimized. Another important issue is to optimize the dwell time for initiate the handover. If the dwell time is a high, then unnecessary delay in the vertical handover decision. If the dwell time is small, then there is high chance of congestion at the hotspot into a particular zone.

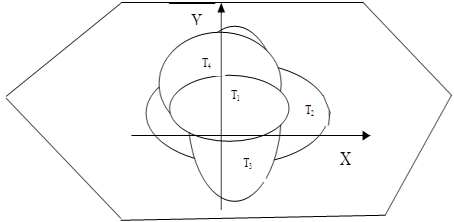

In the hexagonal Cell, There exit n- number of wireless technologies, denoted by $T_i$, Where i = 1.. n.

Figure 1

Let us consider the underlying network support the specific services narrow band and Wide band. The narrow band used for voice call and wide band used for video streaming. We used the parameters to optimize the cost of the vertical handover based on Received signal strength, latency, bandwidth and the covering area.

The cost distribution function expressed as $Z_n = \sum_s \sum_i w_{s,i} . K_{s,i}^n$ where $K_{s,i}^n$ is the cost in the $i^{th}$ parameter to carry out service $s$ on the network $n$. Here $w_{s,i}$ is the 'user-service-priority-weight' assigned to the $i^{th}$ parameter for performing services $s$. We consider the problem as, we will accept the interval for the parameters, where the received signal strength above the considers level and minimizes the cost function.

### III ANALYTIC MODEL

Let $X_i$ ( i=1…k) are present the parameters of all n number of networks such that each network has k numbers of parameters like (latency, received signal strength, covering area). Now $X=X(x_1,…,x_k)$ produce k-dimensional Parametric space. Now $w_{(s,i)}$ is the user service priority weight where $s$ the types of service.

In general if there are $n$ numbers of service types and $m$ numbers of parameters clearly $W_{s,i}$ forming a $m \times n$ matrix. Clearly, $w_{(s,i)} \in R^{\succ 0}$ such that $w_{(s,i)} \in (0,1)$ and $\sum_{i=1}^n w_{s,i} = 1$, for $s =1,2,…m$. So, $W_{s,i}$ generates a stochastic matrix. Here $w_i$ is the continuous random variable defined on the open interval (0,1) such that $\sum_i w_i = 1$. We consider the confidence interval for each of the parameters is defined as ($c_i^j$, $d_i^j$) here j= 1,2,3, presents the networks. The distinct networks have distinctive specification for received signal strength, latency and covering area, represents by the indices, i= 1,2,3…k.

Let $Y_1,…, Y_n$. is a collection of random data with the distribution function $Z(w_1, x), Z(w_2,x),……Z(w_k, x)$. Now $X=(x_1,……,x_k)$ represent the k dimensional parametric space. The joint distribution function L of $w_1,…..w_k$ can be represented as

$$L=Z(w_1,x)Z(w_2,x)…..Z(w_n,x)$$
$$= \prod_{i=1}^n Z(w_i, x) = \prod_{i=1}^n Z(w_i, x = x(x_i)). \quad (1)$$

Now $Z(w_1,x)$, $Z(w_2,x)$, …..$Z(w_n, x)$ are independently and identically distributed. The motivation of the procedure to pick that value for x such that the joint distribution L be optimized for a given set of observed values for $w_1, w_2,….,w_n$.

From the expression (1) we get,

$$\text{Log(L)} = \sum_{i=1}^n Log\{Z(w_i, x = x(x_i))\} \quad (2)$$

From the parametric space, we find out best suitable parameter combination $X = X(\hat{x}'_i,…,\hat{x}'_n)$ such that

$$\left[\frac{\partial L}{\partial x_1}\right]_{X=X(\hat{x}'_i,…\hat{x}'_n)} = 0, \quad \left[\frac{\partial L}{\partial x_2}\right]_{X=X(\hat{x}'_i,…\hat{x}'_n)} = 0,$$
$$\left[\frac{\partial L}{\partial x_3}\right]_{X=X(\hat{x}'_i,…\hat{x}'_n)} = 0 \text{ and } \left[\frac{\partial L}{\partial x_k}\right]_{X=X(\hat{x}'_i,…\hat{x}'_n)} = 0 \quad (3)$$

The above set of equation produce set of k number of nonlinear homogeneous equation such as that
$$AX=b. \quad (4)$$

So we get the solution of the above system as A is the (k, n) dimension coefficient matrix and X is a (n, 1) coefficient matrix. b is a (n,1) matrix.

$$X_j = \sum_{i=1}^k \frac{u_i^T b_j}{\sigma_i} v_i, \text{ for } j = 1,… k. \quad (5)$$

and S= diag{ $\sigma_1, \sigma_2 ….,\sigma_k$ }. Thus $\sigma_i$ are the singular values of matrix A. Here $Sv_i = \sigma_i v_i$, and $S^H u_i = \sigma_i v_i$ now $S^H$ stand for the Hermitian transpose and denotes complex conjugate transpose of a complex matrix.

Now the multiple objective map to the single objective. Here, from the set of vector to the null space, we have to select the vector $X=X(x'_1, x'_2,……, x')$ Where $x'_1$, …., $x'_k$ represent the different parameters related to the Vertical hand over. Since our basic assumption $x_i$ is uniformly distributed over the interval ($c_i^j, d_i^j$) j=1,..,n for the different parameters for the vertical handover, and i= 1..n stands for technologies. Now we estimate the interval ($c_i^j, d_i^j$) for different parameter, j= 1,2 and 3. In a typical mobile radio propagation, the received signal will show fading consisting of very rapid fluctuations at the mean signal level superimposed on relatively slow variations from the mean values. The short term variations are caused by multipath propagation. The amplitude distribution of the signal can be approximated by the Rayleigh distribution. The Rayleigh probability density function used as a model for short term fading in a radio channel. The probability density function p(x) for the amplitude x for the mean power $\lambda$ is

$$p(x) = \frac{x}{\lambda^2} e^{-x^2/2\lambda^2} \quad (6)$$

Now we approximate the interval ($c_i^1, d_i^1$) where j=1 stand for the parameter received signal for the $i^{th}$ technology. Clearly $c_i^1 \leq d_i^1$, is the range. We will consider the range if and only if the 2$^{nd}$ order product derivative is negative for the received signal is maximum. Such that

$$\left[\frac{\partial^2 L}{\partial x^2}\right]_{X=(C_1,\ldots,x_k)} \left[\frac{\partial^2 L}{\partial x^2}\right]_{X=(d_1,\ldots,x_k)} < 0 \quad (7)$$

Since, here we concentrate only the lower level of the signal strength. So the better approximation for strong received signal strength, the required interval is ($c_i^1, +\infty$) with the $\alpha$ degree of rejection. This can be represented by

$$p\{c_i^1 \leq x_1 \prec +\infty\} = 1 - \alpha \quad (8)$$

where p is the probability of acceptance of the interval ($c_i^1, +\infty$)

The parameter $x_1$ follows the distribution function

$$p(x) = \frac{x}{\lambda^2} e^{-x^2/2\lambda^2}. \quad (9)$$

Now, the sample means and variance with respect to the probability density function (9) mean = $\lambda\sqrt{\frac{\pi}{2}} = \mu$ (say)

and variance is $\frac{4-\pi}{2}\lambda^2$, so the stander deviation is $+\lambda\sqrt{\frac{4-\pi}{2}} = \sigma$ (say). If we consider more than (n> 1) number observation from the sample data space, the sample mean and variance are the unbiased estimate of the population mean. Now $A_1, A_2, \ldots, A_n$ are the collected data sample spaces

such that $E(\bar{A}) = E\{(A_1 + A_2 + \cdots + A_n)/n\}$

$$= \frac{\sum_{i=1}^{n} E(A_i)}{n} = \frac{n\mu}{n} = \mu = \lambda\sqrt{\frac{\pi}{2}}$$

and $\text{Var}(\bar{A}) = \frac{\sigma^2}{n} = \lambda^2 \frac{(4-\pi)}{2n}$.

By the Chebyshev's inequality, we get,

$$p\left\{|\bar{A} - \lambda\sqrt{\frac{\pi}{2}}| \geq k\right\} \leq \lambda^2 \frac{(4-\pi)}{2n} \frac{1}{k^2},$$

where k is a positive real number.

Therefore, $p\left\{|\bar{A} - \lambda\sqrt{\frac{\pi}{2}}| \leq k\right\} \to 1$

as $n \to \infty$, true for all $k$, however, small it may be. So we get,

$$\bar{A} - \lambda\sqrt{\frac{\pi}{2}} \leq k \Rightarrow \bar{A} \leq \lambda\sqrt{\frac{\pi}{2}} + k$$

and $-\bar{A} + \lambda\sqrt{\frac{\pi}{2}} \leq k \Rightarrow \bar{A} \geq \lambda\sqrt{\frac{\pi}{2}} - k$

So,

$$\bar{A} \in \left[\lambda\sqrt{\frac{\pi}{2}} - k, \lambda\sqrt{\frac{\pi}{2}} + k\right]. \quad (10)$$

Now we accept our predetermine interval ($c_i^1, d_i^1$) for the better signal strength if and only if the observed value $\bar{A} \in (c_i^1, +\infty)$ with $\alpha$ the degree of rejection according to the expression (8).

We will accept the interval for the parameters $x_1$ which maximized the received signal strength and minimize the cost function.

### IV  SIMULATION RESULT AND DISCUSSION

In the simulation environment, $w_{(s,i)}$ is the user service priority weight where $s$ the type of service is and $i$ is the parameter. All the variables numerically assign a positive integer. The assign values for variable S follows like this, 'below the average' assign a numeric value is 1, for 'accept average' the assign numeric value is 2 and for 'above the average' assign numeric value 3. Now $i$ are the variable indexes for parameters. The Received signal assigns numeric value 4 similarly the assign numeric value 3 for latency, 2 for covering area. The figure 2 and figure 3 represent the received signal strength at the mobile terminal when the mobile terminal cross the one micro cell to new micro cell. The speed of the mobile terminal is uniform and the path the it follows is curved. Figure 2 presents the signal's strength with respect to the time axis. The simulation result presents the received the signal strength within (-5dB, +5dB), when the covering zone getting height priority over other for the higher speed mobile terminal. By changing the priority label, giving the received signal strength highest priority over others. Figure 3 presents the scenario of the received signal strength within (-10dB, 6dB).

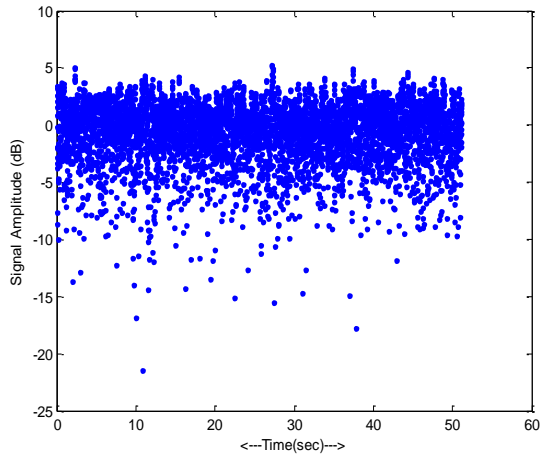

Figure 2

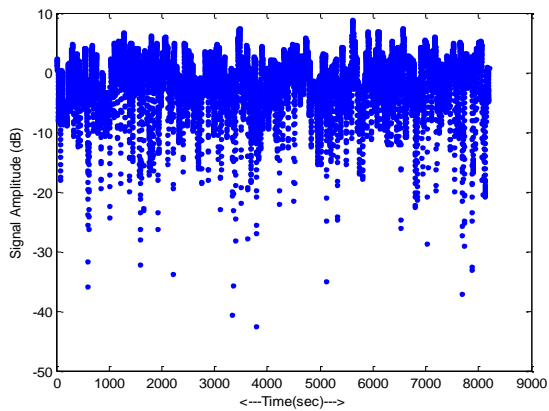

Figure 3

## V CONCLUSION AND FUTURE RESEARCH SECTION

Priority based received signal strength for transporting the video stream over the wireless domain is an open challenge. For the next-generation network, seamless connectivity to the user anywhere at any time very much depends on the strength of the received signal strength. This paper present and indicate the cost effective parametric estimation for the limited case. Lots of research required in this area. This work proposes a cost distribution function to make the soft vertical handover smooth. The extensive research remains with reducing the call dropping probabilities in the actual scenario with respect to real data..